\documentclass[fleqn,usenatbib]{mnras}

\usepackage{newtxtext,newtxmath,xfrac,mathtools,booktabs}

\usepackage[T1]{fontenc}

\DeclareRobustCommand{\VAN}[3]{#2}
\let\VANthebibliography\thebibliography
\def\thebibliography{\DeclareRobustCommand{\VAN}[3]{##3}\VANthebibliography}

\usepackage{graphicx}	
\usepackage{amsmath}	
\usepackage{color}
\usepackage{orcidlink}

\definecolor{Red}{rgb}{0.65,0.08,0.05}
\definecolor{Green}{rgb}{0.05, 0.5, 0.25}
\definecolor{LGreen}{rgb}{0.17, 0.84, 0.57}

\title[ICL and DM halo shapes in CDM and SIDM]{Intracluster light and dark matter halo shapes reflect common assembly, not mutual coupling: shape correspondence in CDM but not SIDM}

\author[G. Martin et al.]{
G. Martin \orcidlink{0000-0003-2939-8668},$^{1}$\thanks{E-mail: garreth.martin@nottingham.ac.uk}
N. A. Hatch \orcidlink{0000-0001-5600-0534},$^{1}$
W. Cui \orcidlink{0000-0002-2113-4863},$^{2,3,4}$
A. Fernandez \orcidlink{0009-0001-4673-526X},$^{1}$
Y. M.~Bah\'{e} \orcidlink{0000-0002-3196-5126},$^{1,5}$
M. S. Fischer \orcidlink{0000-0002-6619-4480},$^{6}$
\newauthor
~M. Montes \orcidlink{0000-0001-7847-0393},$^{7}$
F. R. Pearce \orcidlink{0000-0002-2383-9250},$^{1}$
C. G. Sabiu \orcidlink{0000-0002-5513-5303},$^{8}$
G. Yepes \orcidlink{0000-0001-5031-7936},$^{2,3}$ and
J. Yoo \orcidlink{0000-0002-6841-8329}$^{9}$
\\
$^{1}$School of Physics \& Astronomy, University of Nottingham, University Park, Nottingham NG7 2RD, UK\\
$^{2}$Departamento de F\'{i}sica Te\'{o}rica, M\'{o}dulo 15, Facultad de Ciencias, Universidad Aut\'{o}noma de Madrid, 28049 Madrid, Spain\\
$^{3}$Centro de Investigaci\'{o}n Avanzada en F\'{i}sica Fundamental (CIAFF), Facultad de Ciencias, Universidad Aut\'{o}noma de Madrid, 28049 Madrid, Spain\\
$^{4}$Institute for Astronomy, Royal Observatory, Edinburgh EH9 3HJ, UK\\
$^{5}$Laboratoire d'Astrophysique, \'{E}cole Polytechnique F\'{e}d\'{e}rale de Lausanne (EPFL), Observatoire de Sauverny, 1290 Versoix, Switzerland\\
$^{6}$Donostia International Physics Center (DIPC), Paseo Manuel de Lardizabal 4, 20018 Donostia-San Sebasti\'{a}n, Spain\\
$^{7}$Institute of Space Sciences (ICE, CSIC), Campus UAB, Carrer de Can Magrans, s/n, 08193 Barcelona, Spain\\
$^{8}$Natural Science Research Institute (NSRI), University of Seoul, Seoul 02504, Republic of Korea\\
$^{9}$Korea Astronomy and Space Science Institute (KASI), Daedeokdae-ro, Daejeon 34055, Republic of Korea
}

\date{Accepted XXX. Received YYY; in original form ZZZ}

\pubyear{\the\year{}}

\begin{document}
\label{firstpage}
\pagerange{\pageref{firstpage}--\pageref{lastpage}}
\maketitle

\begin{abstract}
The intracluster light (ICL) has been proposed as a luminous tracer of the dark matter (DM) halo shape in galaxy clusters, with recent observational studies and $\Lambda$CDM simulations suggesting close alignment between the two components. We test whether this correspondence holds in a self-interacting dark matter (SIDM) cosmology using matched CDM and SIDM ($\sigma_\mathrm{DM}/m = 0.1$ and $0.5\;\mathrm{cm^{2}\,g^{-1}}$) hydrodynamical cluster simulations from \textsc{TheThreeHundred} project. We measure 3D axis ratios and major-axis orientations as a function of radius for both the DM halo and the stellar (BCG+ICL) distribution, and quantify the spatial correspondence using the Weighted Overlap Coefficient. In the SIDM simulations, DM haloes become significantly rounder in their inner regions with increasing DM cross-section, while the stellar distribution retains the same elongated morphology in both CDM and SIDM. This decoupling is robust across two baryonic physics models and two resolution levels, and is detected at $>5\sigma$ for $\sigma_\mathrm{DM}/m = 0.5\;\mathrm{cm^{2}\,g^{-1}}$. It arises because the stellar shape is governed by the accretion geometry of tidally stripped satellites along the large-scale structure, which is identical in CDM and SIDM, whereas, in SIDM, halo shapes respond additionally to local self-interactions. The ICL--DM shape correspondence demonstrated in CDM therefore breaks down in an SIDM universe. 
\end{abstract}



\section{Introduction}
\label{sec:intro}

The shapes of dark matter (DM) haloes encode information about both the assembly history of cosmic structure and the fundamental nature of DM. In the standard cold dark matter (CDM) framework, haloes are generically triaxial, with a preference for prolate configurations that reflects anisotropic infall along the filaments of the cosmic web \citep[e.g.][]{Dubinski1991, Jing2002, Allgood2006, Despali2014}. The inclusion of baryonic physics tends to make haloes rounder in their inner regions, where the central galaxy dominates the gravitational potential \citep[e.g.][]{Kazantzidis2004, Debattista2008, Chua2022}. In self-interacting dark matter (SIDM) models, momentum exchange between DM particles provides an additional isotropisation mechanism that produces significantly rounder inner haloes than CDM at all mass scales \citep{Dave2001, Peter2013, Brinckmann2018, Despali2022, Giocoli2026}.

The sensitivity of the halo shape to DM microphysics has motivated efforts to identify luminous tracers that can be used to map the DM morphology. The intracluster light (ICL), composed of stars tidally stripped from satellite galaxies during cluster assembly \citep[e.g.][]{Byrd1990, Murante2007, Contini2014, Mihos2017, Montes2022, Brown2024, Contreras-Santos2025}, has emerged as one promising candidate. Unlike, for example, the hot intracluster medium, whose distribution is shaped by gas-dynamical processes such as shocks and AGN feedback, intracluster stars are effectively collisionless and their dynamics are governed primarily by the cluster gravitational potential. Both observational \citep{Montes2019b, Cha2025, Ellien2025} and simulation-based \citep{Alonso-Asensio2020, Gonzalez2021, Yoo2022, Yoo2024, Yoo2025, Contreras-Santos2024, Butler2025, Kimmig2025, Fernandez2026} studies have demonstrated that the large-scale morphology of the ICL corresponds closely to the total mass distribution in CDM cosmologies. \citet{Fernandez2026}, using the Hydrangea simulations, found close agreement between the 3D axis ratios and orientations of the ICL and DM distributions.

It is important, however, to distinguish between two senses in which the ICL might ``trace'' DM. Individual intracluster stars orbit in the cluster potential as collisionless particles, but even at this level they do not sample the DM phase-space uniformly. Stripped stars occupy systematically lower specific energies and more radial orbits than the DM \citep{Butler2025, Martin2026}, which is a consequence of the differential stripping of stars and DM from infalling satellites \citep{Smith2016, Martin2026}. The \textit{shape of the stellar distribution} is a further step removed: it reflects not the orbits of individual stars but the spatial pattern in which stripped material was deposited during the cluster's assembly. This depends on the directions and timing of satellite accretion, which are set by the large-scale structure and the satellite mass function \citep{Brown2024, Martin2026}.

The close correspondence between ICL and DM shapes reported by these studies has been established mainly in CDM cosmologies. The first comparison in an SIDM context was recently presented by \citet{Yoo2026}, who applied the Weighted Overlap Coefficient \citep[WOC;][]{Yoo2022} to two C-EAGLE clusters re-simulated with $\sigma_\mathrm{DM}/m = 1\;\mathrm{cm^{2}\,g^{-1}}$. They found that the BCG+ICL remains the component whose shape most closely matches that of the DM in both CDM and SIDM, but that in SIDM the correspondence between the gas and DM distributions is stronger than in CDM, reflecting the effectively collisional nature of both gas and DM self-interactions. In this study we explore whether ICL shapes trace the DM halo in an SIDM universe, using matched CDM and SIDM hydrodynamical cluster simulations from \textsc{TheThreeHundred} project \citep{Cui2018}. We compare five simulation configurations that span two DM models, two baryonic physics prescriptions, and two resolution levels. This represents the first systematic comparison of ICL and DM shapes in SIDM using a statistical sample of clusters with full hydrodynamical baryonic physics.

\section{Simulations}
\label{sec:sims}

All simulations are drawn from \textsc{TheThreeHundred} project, which comprises zoom-in re-simulations of 324 massive galaxy clusters ($M_{200} \gtrsim 8 \times 10^{14}\,h^{-1}\,\mathrm{M_\odot}$ at $z=0$) selected from the MultiDark Planck~2 (MDPL2) $N$-body simulation \citep{Klypin2016}. All runs share the same initial conditions and cosmological parameters, adopting a $\Lambda$CDM cosmology consistent with \citet{Planck2016}, with $(\Omega_{\rm m}, \Omega_\Lambda, \Omega_{\rm b}, \sigma_8, n_s, h) = (0.307, 0.693, 0.048, 0.823, 0.96, 0.678)$. The simulations are run at two resolution levels (3k and 7k), summarised in Table~\ref{tab:simulations}.

The \textsc{Gadget-X} CDM runs \citep{Rasia2015} use a modernised SPH solver based on \textsc{Gadget-3} \citep{Beck2016} with thermal AGN feedback \citep{Steinborn2015} and the multiphase star formation model of \citet{Springel2003}, and are available at the 3k resolution only. The \textsc{Gizmo-Simba} runs \citep{Cui2022} employ the meshless finite-mass hydrodynamics solver in \textsc{Gizmo} \citep{Hopkins2015} with the \textsc{Simba-C} galaxy formation model \citep{Hough2023}, which includes kinetic jet-mode AGN feedback, and are available at both 3k and 7k resolutions.

The SIDM variants use the same \textsc{Gizmo-Simba} baryonic physics and initial conditions as the CDM runs, adding velocity-independent elastic scattering with total cross-sections per unit mass of $\sigma_\mathrm{DM}/m = 0.1$ and $0.5\;\mathrm{cm^{2}\,g^{-1}}$ at both resolution levels. These cross-sections bracket the range of current observational constraints on cluster scales \citep[e.g.][]{RobertsonA2019, Harvey2019, Sagunski2021, Eckert2022, Harvey2025, Adhikari2025}. The SIDM runs are available for the 50 most massive clusters in the sample, with $z = 0$ halo masses $10^{14.9} \lesssim M_{200}/\mathrm{M_\odot} \lesssim 10^{15.4}$, and we restrict our analysis to these throughout to enable a like-for-like comparison across all five simulation configurations and where SIDM effects on halo structure are most readily constrained observationally \citep[e.g.][]{Sagunski2021, Eckert2022}.

\begin{table}
\caption{Summary of the simulation suites used in this work. $m_\mathrm{DM}$, $m_\mathrm{gas}$, and $m_\star$ are the DM, gas, and stellar particle masses. $\epsilon$ is the Plummer-equivalent gravitational softening length for collisionless particles (comoving / physical, whichever is smaller; the 3k runs use fixed comoving softening).}
\label{tab:simulations}
\setlength{\tabcolsep}{7pt}
\begin{tabular}{lcccc}
\hline
 & \textsc{Gadget-X} & \multicolumn{3}{c}{\textsc{Gizmo-Simba}} \\
 & CDM & CDM & \multicolumn{2}{c}{\hfill SIDM\hfill\mbox{}} \\
\hline
$\sigma_{\rm DM}/m$ [$\mathrm{cm^2\,g^{-1}}$] & 0 & 0 & \hfill 0.1\hfill & \hfill 0.5\hfill \\
Resolution & 3k & \multicolumn{3}{c}{3k / 7k} \\
Code & \textsc{Gadget-3} & \multicolumn{3}{c}{\textsc{Gizmo}} \\
Hydro. method & SPH & \multicolumn{3}{c}{MFM} \\
Subgrid & \citet{Rasia2015} & \multicolumn{3}{c}{\citet{Hough2023}} \\
\hline
 & \multicolumn{2}{c}{\textit{3k}} & \multicolumn{2}{c}{\textit{7k}} \\
\hline
$m_\mathrm{DM}$ [$\mathrm{M_\odot}$] & \multicolumn{2}{c}{$2\times10^{9}$} & \multicolumn{2}{c}{$2\times10^{8}$} \\
$m_\mathrm{gas}$ [$\mathrm{M_\odot}$] & \multicolumn{2}{c}{$3\times10^{8}$} & \multicolumn{2}{c}{$4\times10^{7}$} \\
$m_\star$ [$\mathrm{M_\odot}$] & \multicolumn{2}{c}{$1.5 - 3\times10^{8}$} & \multicolumn{2}{c}{$2-4\times10^{7}$} \\
$\epsilon$ [kpc] & \multicolumn{2}{c}{7.4} & \multicolumn{2}{c}{3.7 / 1.8} \\
\hline
\end{tabular}
\end{table}

\section{Method}
\label{sec:method}

\subsection{Shape tensor analysis}
\label{sec:shape}

We compute the 3D shape of the DM halo and stellar (BCG+ICL) distributions separately as a function of radius. For each cluster at each snapshot, we first excise all particles identified by the Amiga Halo Finder \citep[\textsc{ahf};][]{Knollmann2009}   as belonging to satellite subhaloes, then determine independent centres for the DM and stellar components using a shrinking-sphere algorithm \citep{Power2003}. Starting from the \textsc{AHF}-identified cluster centre with an initial radius of $r_{200}$, we iteratively compute the centre-of-mass, recentre, and reduce the radius by a factor of 0.99 until fewer than 100 particles remain.

We then compute the mass-weighted inertia tensor in radial shells, using the respective converged shrinking-sphere centre of each component. The outer shell boundary steps inward from $r_{200}$, with the standard shell at step $n$ spanning $[r_{n+1},\, r_n]$, where $r_{n+1} = 0.99\,r_n$. If a shell contains fewer than 5000 particles, the inner boundary is moved inward until the shell encloses exactly 5000 particles, ensuring a stable tensor measurement while allowing the outer radius to continue stepping uniformly. At each shell we apply the iterative ellipsoidal refinement of the unweighted inertia tensor \citep[e.g.][]{Dubinski1991, Allgood2006}. The initial spherical shell defines a starting ellipsoid, whose axis ratios and orientation are updated iteratively until convergence, following the procedure described in \citet{Fernandez2026}. The eigenvalues of the converged tensor yield the axis ratios $b/a$ (intermediate-to-major) and $c/a$ (minor-to-major) and the eigenvectors define the position angle (PA) of the principal axes.

Using synthetic triaxial gNFW distributions \citep{Wyithe2001} sampled at the typical 3k particle count, scale radius and $r_{200}$, we recover axis ratios to within $\lesssim 2$ per cent at all radii considered in this work, even for a cored inner profile corresponding to an inner slope of $\gamma = 0$. Applying the \citet{Fischer2023} shape sensitivity criterion $\xi$ directly to the simulation data, we also find median $\xi > 1$ at all radii for all models, confirming that density gradients are sufficiently steep relative to Poisson noise for the iterative shape tensor to yield reliable axis ratio measurements.

\subsection{Weighted Overlap Coefficient}
\label{sec:woc}

To complement the shape tensor analysis, we compute the Weighted Overlap Coefficient (WOC; \citealt{Yoo2022}) in 3D, which quantifies the morphological correspondence between two density fields by computing the fractional overlap of iso-density surface matched by enclosed volume, weighted by enclosed volume and threshold density, and normalised to the range $[0,\,1]$. Unlike the shape tensor, the WOC does not assume ellipsoidal symmetry, providing a complementary diagnostic. For each cluster at each snapshot, we construct 3D mass-weighted density volumes for the DM and stellar (BCG+ICL) components on a regular cubic grid, with satellite particles excised as in Section~\ref{sec:shape}. At each target radius $r$, the grid is centred on the stellar shrinking-sphere centre with half-width $1.5\,r$ and $25^3$ voxels, so that the voxel size scales with the radius being probed, with a floor at the gravitational softening length. We evaluate the WOC at 12 logarithmically spaced radii from $r/r_{200} = 0.02$ to $1.0$ using \textsc{pywoc}\footnote{\url{https://github.com/csabiu/WOC}} \citep{Yoo2022,Yoo2024}. Our approach ensures that the spatial sampling is appropriate at each scale, avoiding the shot-noise bias that would arise from evaluating a single fine grid across the full radial range. We validate the method using synthetic co-aligned NFW (DM) and Hernquist (stellar) profiles sampled at the 3k particle count, with the DM halo using the same NFW profile ($\gamma = 1$), scale radius and $r_{200}$ as the tests described in Section \ref{sec:shape}; because these distributions are perfectly co-centred and spherically symmetric the true WOC is unity by construction, and we recover $\mathrm{WOC} \gtrsim 0.95$ at all radii, confirming that the gridding and smoothing do not introduce significant systematic bias.

We stack results from our shape tensor and WOC analysis over all 50 clusters common to all five simulation suites and over ten snapshots spanning $z \approx 0.2$--$0$ (snapshots 119--128), chosen to improve statistical robustness and reduce snapshot-to-snapshot variance. We then compute the median and 16th--84th percentile range at each radius.

\section{Results}
\label{sec:results}

\subsection{Shape and orientation}
\label{sec:results_3d}

\begin{figure*}
    \centering
    \includegraphics[width=0.95\textwidth]{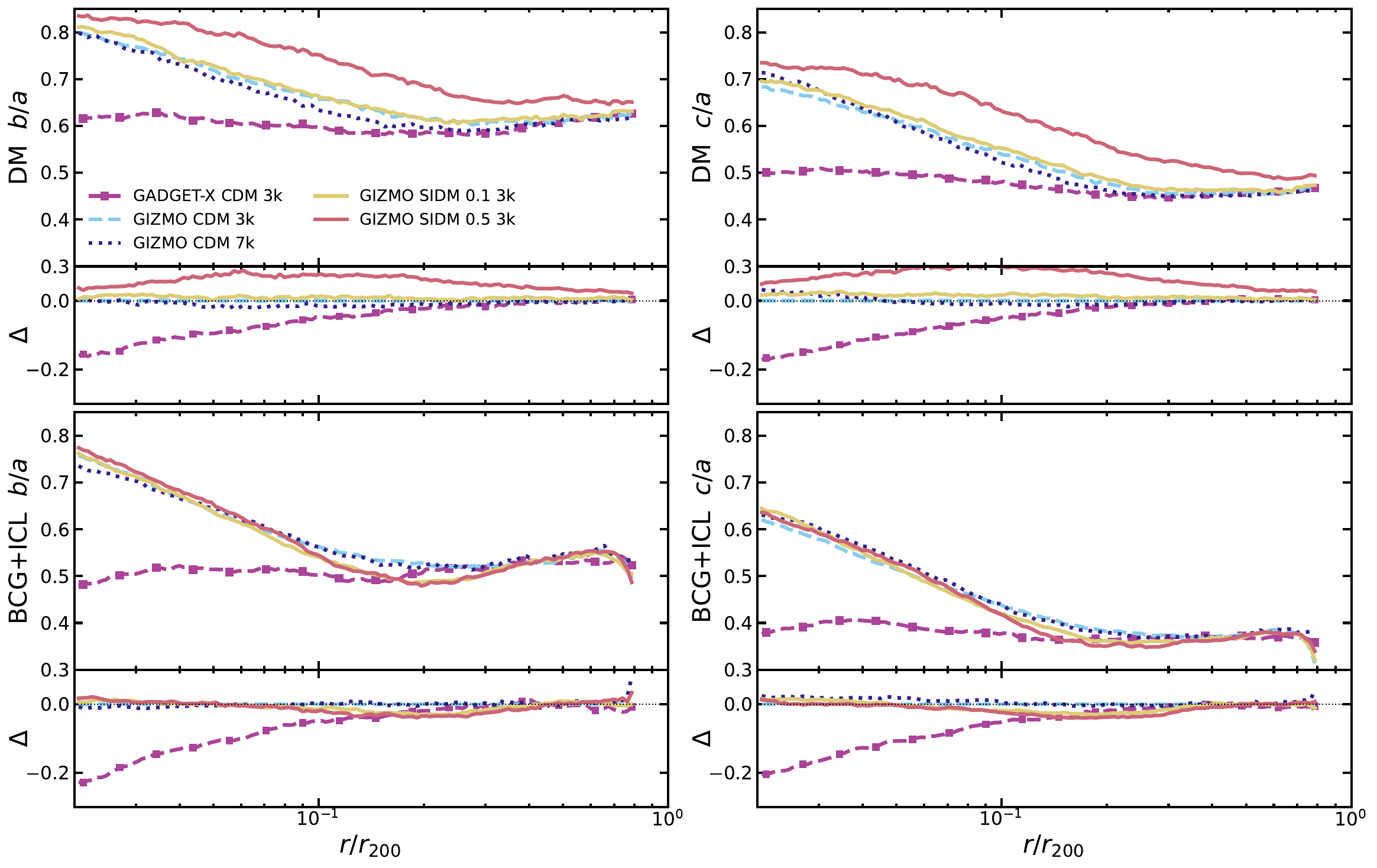}
    \caption{Stacked radial profiles of the 3D axis ratios $b/a$ (left column) and $c/a$ (right column) for the DM halo (top) and stars (BCG+ICL; bottom). Each panel is split into sub-panels showing the median profile above and the median per-cluster-pair residual, $\Delta$, relative to the \textsc{Gizmo-Simba} CDM 3k model below. The typical cluster-to-cluster scatter in the DM and stellar axis ratios, measured as half the 16th--84th percentile range, is 0.1--0.15.}
    \label{fig:shape_profiles}
\end{figure*}

Figure~\ref{fig:shape_profiles} shows the stacked radial profiles of the DM and stellar axis ratios $b/a$ and $c/a$ for all five simulation models. At large radii ($r \gtrsim 0.2\,r_{200}$) the DM axis ratios are similar across all models, but towards the centre the CDM and SIDM runs diverge. The \textsc{Gizmo-Simba} CDM configurations all produce DM axis ratios of $0.7 \lesssim b/a \lesssim 0.8$ and $0.6 \lesssim c/a \lesssim 0.7$ at $r < 0.1\,r_{200}$, with the \textsc{Gadget-X} run producing more elongated inner haloes with $b/a \approx 0.6$ and $c/a \approx 0.5$. As expected from collisional isotropisation \citep[e.g.][]{Brinckmann2018, RobertsonA2019, Banerjee2020}, the SIDM models are systematically rounder, especially within the inner $\sim 0.1\,r_{200}$ and for the $\sigma_\mathrm{DM}/m = 0.5\;\mathrm{cm^{2}\,g^{-1}}$ model.

The DM halo is systematically rounder than the stellar distribution in all five models, including in CDM. This likely reflects their different assembly channels: the DM halo is built from smooth accretion \citep{Genel2010} and a wide range of stripped satellite haloes, sampling a relatively isotropic distribution of infall directions, while the ICL is dominated by a smaller number of massive progenitors that tend to produce a more elongated distribution \citep[see also][]{Brown2024, Butler2025, Fernandez2026, Martin2026}.

Comparing the two baryonic physics models at matched 3k resolution reveals that the choice of subgrid physics has a considerable effect on both the stellar and DM shapes within $r \lesssim 0.1\,r_{200}$, where the \textsc{Gadget-X} and \textsc{Gizmo-Simba} CDM runs produce visibly different axis ratio profiles for both components. This likely reflects differences in how the BCG is assembled in the two models. The treatment of AGN feedback and star formation affects the stellar masses and structure of the most massive satellites, altering the stellar debris they deposit in the inner cluster. It also changes the depth of the baryon-dominated potential well at small radii, which in turn influences the DM shape through its gravitational response to the central stellar mass. By contrast, changing the DM model from CDM to SIDM at fixed baryonic physics produces little discernible change in the stellar shape, despite the substantial rounding of the DM halo: the BCG+ICL axis ratio profiles remain very close to those of the corresponding CDM run. Figure~\ref{fig:shape_divergence} quantifies the resulting DM--stellar mismatch as a function of radius.

The left panel of Figure~\ref{fig:shape_divergence} shows this offset directly as the difference $\Delta(b/a)$ between the DM and stellar axis ratios. In the \textsc{Gizmo-Simba} CDM runs, $\Delta(b/a)$ is small at all radii and decreases towards the centre, where the DM and stellar mass are both deposited by a small number of major merger events \citep{Brown2024, Martin2026}, producing similar shapes. The \textsc{Gadget-X} CDM run shows a larger $\Delta(b/a)$ in the centre, indicating that the baseline DM--stellar shape offset is itself sensitive to the choice of baryonic physics model \citep[e.g. see][]{Fernandez2026}; the residual panels in Figure~\ref{fig:shape_profiles} confirm that this sensitivity is confined to the inner $r \lesssim 0.1\,r_{200}$. On the other hand, the SIDM models depart significantly from the CDM reference \textsc{Gizmo-Simba} simulation only in the DM, with the stellar residuals remaining close to zero at all radii and the DM becoming rounder such that $\Delta(b/a)$ rises steeply in excess of the CDM values for the same radii. A complementary signature appears in the orientation of the major axis shown in the central panel. The DM--stellar PA offset rises from $\sim 6$--$8\degr$ in CDM to $\sim 10$--$11\degr$ in the SIDM models at $0.1\,r_{200}$.

Splitting the sample at the median formation redshift ($z_{50} = 0.47$) reveals that the DM--stellar axis ratio offset is larger for early-forming clusters in all models, but the difference grows substantially with cross-section: $\Delta(\mathrm{early} - \mathrm{late}) \approx 0.01$ in CDM versus $\approx 0.08$ for $\sigma_\mathrm{DM}/m = 0.5\;\mathrm{cm^{2}\,g^{-1}}$, consistent with cumulative DM scattering progressively rounding the inner halo while the collisionless stellar distribution retains its original shape.

\subsection{Weighted Overlap Coefficient}
\label{sec:results_woc}

\begin{figure*}
    \centering
    \includegraphics[width=0.95\textwidth]{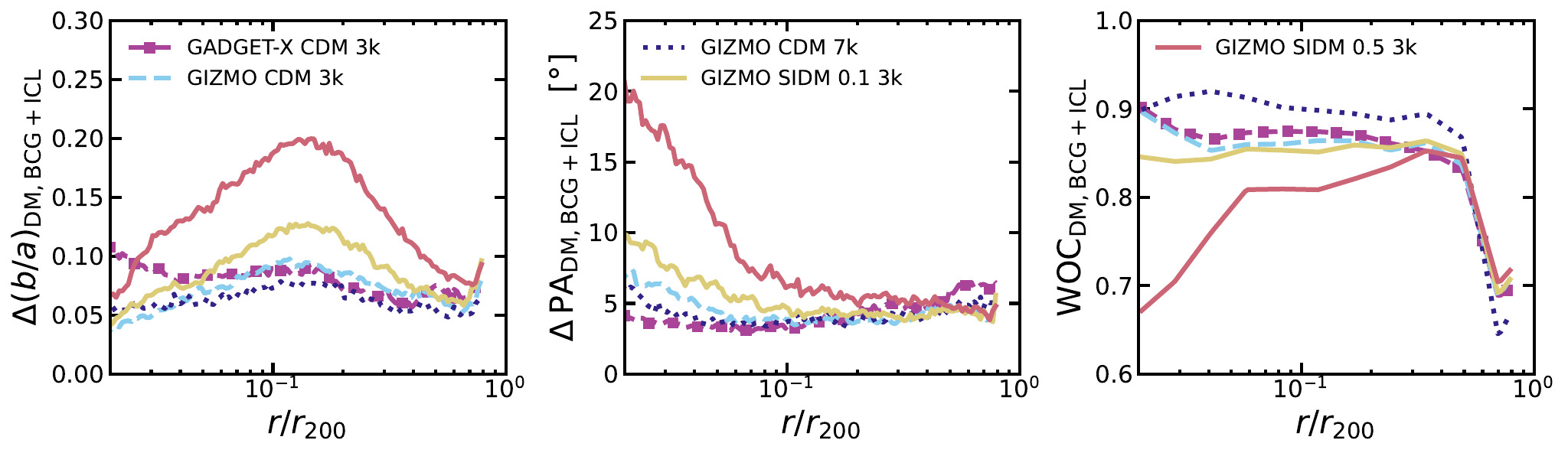}
    \caption{Stacked radial profiles of the DM--stellar axis ratio difference $\Delta(b/a)$ (left), defined as the DM axis ratio minus the stellar axis ratio at matched radius, the DM--stellar major-axis position-angle misalignment $\Delta\mathrm{PA}$ (centre), and the 3D WOC (right) between the DM and stellar (BCG+ICL) density fields.}
    \label{fig:shape_divergence}
\end{figure*}

The bottom-right panel of Figure~\ref{fig:shape_divergence} shows the 3D WOC between the DM and stellar (BCG+ICL) density fields as a function of radius for each of the simulation models. At large radii ($r \gtrsim 0.3\,r_{200}$), all models maintain high WOC values ($\gtrsim 0.85$), indicating good overall morphological correspondence between the DM and stellar distributions. The CDM models maintain relatively high WOC values at all radii, whereas the SIDM models show a progressive decline towards the centre, with the $\sigma_{\mathrm{DM}}/m = 0.5\,\mathrm{cm^2\,g^{-1}}$ model showing the largest decrease. The WOC is systematically higher at 7k resolution, at least partly because noisier density fields at lower particle counts raise the resolution-dependent floor identified in our validation tests (Section~\ref{sec:woc}), but this offset is smaller than the CDM--SIDM difference at matched resolution.

Unlike the axis ratio differences, which peak at $r \sim 0.1\,r_{200}$, the WOC continues to decline towards the centre rather than recovering, suggesting that the morphological mismatch between the DM and stellar density fields in SIDM extends into the BCG-dominated regime even where the axis ratios of the two components converge. This likely reflects the sensitivity of the WOC to aspects of the density field beyond ellipsoidal shape, such as the increased major-axis misalignment seen in the central panel of Figure~\ref{fig:shape_divergence}, or centroid offsets between the two components. In particular, the WOC computes the overlap of 
volume-matched iso-density surfaces on a common grid, so the oscillation of the BCG relative to the DM potential minimum \citep[e.g.][]{Kim2017,Fischer2023b}, for which our shrinking-sphere estimates yield average offsets exceeding 10\,kpc in the $\sigma_{\mathrm{DM}}/m = 0.5\,\mathrm{cm^2\,g^{-1}}$ model, will reduce the WOC at small radii without necessarily affecting the axis ratios, which are measured in their respective frames.

\subsection{Pairwise model comparison}

The preceding sections establish that the CDM and SIDM models produce ICL shapes that are decoupled from the shape of the DM halo, resulting in clear DM--stellar shape offsets in the SIDM models. To quantify the statistical significance of these differences, we compare the three 3k simulation models at $r = 0.1\,r_{200}$ using a two-level hierarchical Bayesian model. The first level describes how individual snapshot measurements scatter around a cluster mean; the second level describes how those cluster means are distributed across the population, characterised by a population mean $\mu$ and cluster-to-cluster scatter $\tau$. This structure prevents the repeated snapshots per cluster from being treated as independent data points, which would otherwise artificially inflate statistical power. Pairwise significance between models is computed as the posterior $z$-score of the difference in population means, $\langle \Delta\mu \rangle / \sigma(\Delta\mu)$.

\begin{figure}
\centering
\includegraphics[width=0.45\textwidth]{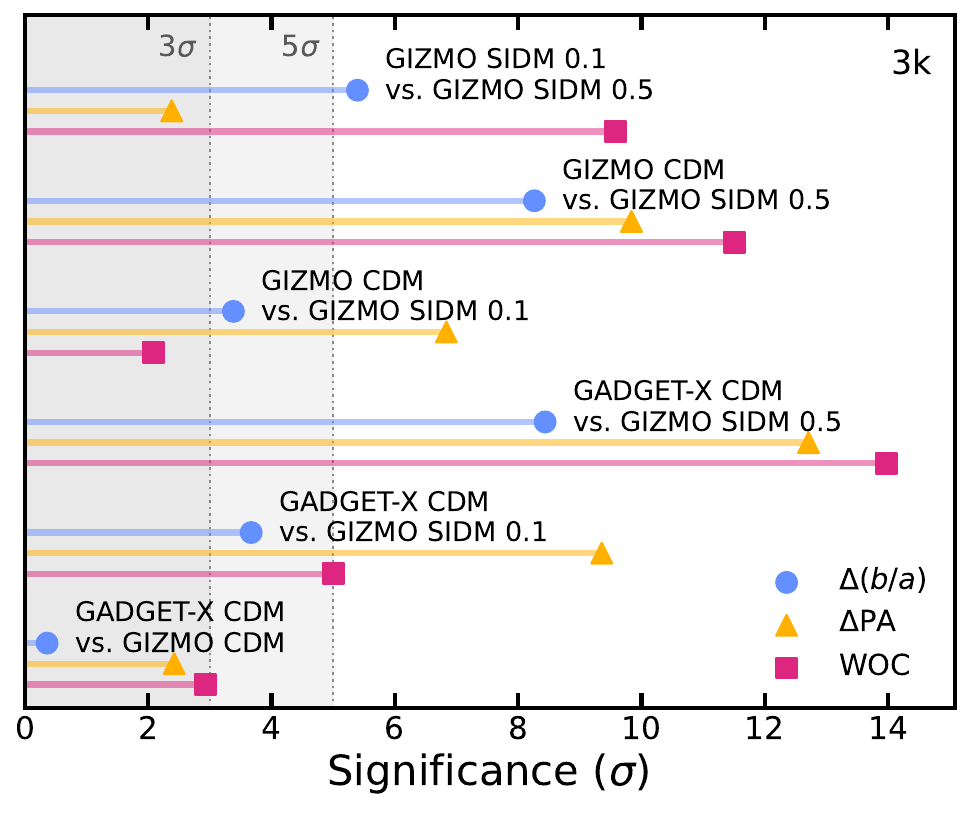}
\caption{Pairwise significance of differences between all 3k models at $r = 0.1\,r_{200}$, estimated from a two-level hierarchical Bayesian model. Circles, triangles, and squares show the posterior $z$-score for $\Delta(b/a)$, $\Delta\mathrm{PA}$, and the WOC, respectively. Shaded bands mark the $<3\sigma$ and $<5\sigma$ regions.}
\label{fig:significance}
\end{figure}

Figure~\ref{fig:significance} shows the resulting pairwise $z$-scores. The two CDM models (\textsc{Gizmo-Simba} and \textsc{Gadget-X}) are mutually consistent at better than $3\sigma$ for $\Delta(b/a)$, $\Delta\mathrm{PA}$ and WOC. The $\sigma_\mathrm{DM}/m = 0.5\;\mathrm{cm^{2}\,g^{-1}}$ model differs from every CDM model by more than $5\sigma$ in all three diagnostics. The $\sigma_\mathrm{DM}/m = 0.1\;\mathrm{cm^{2}\,g^{-1}}$ model exceeds $3\sigma$ against all CDM models in both $\Delta(b/a)$ and $\Delta\mathrm{PA}$, and exceeds $5\sigma$ against \textsc{Gadget-X} CDM in WOC, but falls below $3\sigma$ against \textsc{Gizmo-Simba} CDM in WOC, leaving that comparison marginal. Taken together, the decoupling between DM and stellar morphology is unambiguously detected for $\sigma_\mathrm{DM}/m = 0.5\;\mathrm{cm^{2}\,g^{-1}}$ and remains significant for $\sigma_\mathrm{DM}/m = 0.1\;\mathrm{cm^{2}\,g^{-1}}$ in $\Delta(b/a)$ and $\Delta\mathrm{PA}$. We evaluate the comparison at $r = 0.1\,r_{200}$ ($\sim 200\,$kpc in these clusters) because this is roughly where the axis ratio divergence peaks; the WOC signal would strengthen at smaller radii. Strong lensing can constrain the projected total mass morphology well at these radii, but disentangling the DM shape from the baryonic contribution remains challenging, particularly given the sensitivity of both components to subgrid physics at these scales.

The correspondence between ICL and DM shapes in CDM has sometimes been interpreted as the stellar distribution tracing the cluster potential, but our results suggest a different origin: both components inherit their shape from the same accretion geometry \citep{Allgood2006, Despali2014, Fernandez2026} and their agreement is therefore a consequence of shared infall directions rather than dynamical equilibrium in a common potential. SIDM modifies the inner DM shape through collisional isotropisation, but produces comparatively little change in the stellar morphology, which retains the imprint of the cluster assembly history. We test this picture by computing the Spearman rank correlation of per-cluster median PA misalignment between all pairs of 3k models. A lack of correlation in the PA misalignment between CDM and SIDM models would indicate that SIDM erases the imprint of each cluster's accretion history, such that clusters that are highly aligned in CDM are not preferentially highly aligned in SIDM. All six correlations are positive and significant ($r_s = 0.52$--$0.64$), with CDM--SIDM correlations comparable to CDM--CDM ones, consistent with the rank ordering of PA misalignment being set by each cluster's individual assembly history while self-interactions impose a roughly uniform shift on the population mean without reordering which clusters are most misaligned.

\section{Discussion}
\label{sec:discussion}

The central result of this work is that SIDM-induced changes to the DM halo shape are not accompanied by corresponding changes in the stellar (BCG+ICL) morphology. Stripped stars remain collisionless and retain this imprint regardless of the DM model. These results are robust to numerical resolution, with the \textsc{Gizmo-Simba} CDM 3k and 7k runs well converged at all radii considered.

Although the individual orbits of stripped stars are themselves biased relative to DM, occupying lower specific energies and more radial orbits \citep{Butler2025, Martin2026}, the \textit{shape} of the collective stellar distribution is set by the accretion geometry that deposited the material in the first place. The presence of an SIDM core, formed by collisional isotropisation of DM orbits \citep{Peter2013, Despali2022}, may change where along a satellite's orbit its stars are liberated, but not the direction from which the satellite arrives, and the DM axis ratios differ only marginally between models outside the innermost regions (Figure~\ref{fig:shape_profiles}). In the innermost regions, where the SIDM core has its strongest effect on the potential shape, the stellar mass is in any case dominated by the BCG, whose structure is set by its own formation history rather than the current DM profile shape. This is consistent with the sensitivity of the inner shapes to baryonic physics seen in Figure~\ref{fig:shape_profiles}. The choice of subgrid model affects how the BCG is assembled and therefore alters the inner shapes of both components, while changing the DM model at fixed baryonic physics does not.

This picture assumes that the stripping and identification of satellites do not themselves differ between DM models in ways that bias our measurements. Self-interactions alter the central density profile of the host \citep{Kaplinghat2016}, stripping efficiency depends on the inner density slope of the satellite \citep{Penarrubia2010, Martin2024}, SIDM suppresses the survival of self-bound subhaloes \citep{Turner2021}, and subhalo finders including \textsc{ahf} are less reliable at recovering a subhalo's outer particles \citep{Onions2012}. Each of these mechanisms alters how much stellar mass a satellite loses and where along its orbit, but not the direction from which it arrives, which is set by the large-scale structure and is essentially identical between our CDM and SIDM runs; it is this accretion direction, rather than stripping efficiency, that sets the ICL geometry.

Consistent with this, the stellar axis ratio remains stable across models at fixed baryonic physics (Figure~\ref{fig:shape_profiles}), while we measure only a mild redistribution in the stacked stellar density profile: a deficit in the innermost region and an excess peaking at $r \sim 0.15$--$0.2\,r_{200}$, reaching $\sim 20$ ($\sim 50$) per cent for $\sigma_\mathrm{DM}/m = 0.1$ ($0.5$)\,$\mathrm{cm^2\,g^{-1}}$, comparable to the offsets induced by resolution and by the choice of baryonic physics model respectively. This is the pattern expected if satellites in SIDM deposit their stars earlier along their orbits, and it is not accompanied by any change in the stellar axis ratio. A full characterisation of the ICL mass budget and its radial distribution is left to a forthcoming paper (Martin et al. in prep.).

These results have direct implications for ICL-based shape measurements. \citet{Fernandez2026} demonstrated that the ICL and DM shapes agree to within $\Delta(b/a) \approx 0.07$ in the Hydrangea CDM simulations, indicating a close correspondence between the two components with a small but non-zero misalignment, and proposed that the ICL could serve as a robust proxy for the DM morphology. In CDM we find the same close correspondence, because both components inherit their shape from the same accretion geometry rather than the ICL tracking the DM potential, and it is this shared origin that allows the correspondence to break down once self-interactions reshape the DM alone. \citet{Yoo2026}, using the projected WOC applied to two C-EAGLE clusters re-simulated with $\sigma_{\mathrm{DM}}/m = 1\,\mathrm{cm^2\,g^{-1}}$, found that the BCG+ICL remains the closest morphological match to the projected DM distribution, even when DM is self-interacting, though with a modestly reduced correspondence in SIDM. Using an independent simulation suite and lower cross-sections ($\sigma_{\mathrm{DM}}/m = 0.1$ and $0.5\,\mathrm{cm^2\,g^{-1}}$), our results are consistent with this at large radii, where the 3D WOC remains high in all models, but reveal that the correspondence breaks down in the inner halo in SIDM: the DM becomes rounder while the stellar distribution does not, producing a growing axis ratio difference that peaks at $r \approx 0.1\,r_{200}$, an increasing major-axis misalignment, and a WOC that declines monotonically towards the centre. The projected WOC, integrated over the full cluster extent, is dominated by the outer regions where the CDM--SIDM difference is small, which explains the modest signal found by \citet{Yoo2026}; the radially resolved 3D diagnostics presented here reveal a strong decoupling concentrated within the inner halo.

We have considered only velocity-independent cross-sections, for which current cluster-scale limits span $\sigma_\mathrm{DM}/m \lesssim 0.2-0.4\,\mathrm{cm^2\,g^{-1}}$ depending on the probe \citep{Harvey2019, Sagunski2021, Eckert2022}, placing our larger cross-section at the upper end of the allowed range. However, the tension between these limits and the larger cross-sections required to produce cores on galactic scales increasingly favours models in which the cross-section falls with relative velocity \citep[e.g.][]{Kaplinghat2016, Eckert2022}, and recent simulations implement such models directly \citep{RobertsonA2019,Ragagnin2024, Despali2025}. The decoupling arises simply because self-interactions act on the DM while the stars remain collisionless. Its strength is therefore governed principally by the effective scattering rate at cluster velocities rather than by the detailed form of the cross-section, so a velocity-dependent model should produce a shape decoupling bracketed by our $\sigma_\mathrm{DM}/m = 0.1$ and $0.5\,\mathrm{cm^2\,g^{-1}}$ results, according to its effective cross-section at cluster velocity scales.

In principle, a measured mismatch between the ICL shape and an independent constraint on the total mass morphology from gravitational lensing would constitute evidence for a departure from CDM physics. The strongest shape divergence occurs at $r \sim 0.1\,r_{200}$ ($\sim 200\,$kpc in these clusters), where strong lensing can constrain the projected mass morphology well, though since lensing measures total rather than DM mass, 
interpreting any mismatch with the ICL shape in terms of DM microphysics would require modelling the baryonic contribution. At larger radii, stacked weak lensing can measure halo ellipticities \citep{Shin2018, Gonzalez2024}, but in this regime CDM--SIDM shape differences are small \citep{Robertson2023}. New wide-field surveys with LSST and \textit{Euclid} will deliver both the deep photometry needed to measure the ICL shape and the lensing constraints needed to map the projected total mass distribution for statistical samples of clusters. Detecting the signal predicted here will require combining these datasets at radii where both the ICL and total mass morphologies can be reliably measured, along with forward-modelling to account for the contribution of baryons to the projected mass distribution.

\section{Conclusions}
\label{sec:conclusions}

Using matched CDM and SIDM hydrodynamical simulations of massive galaxy clusters from \textsc{TheThreeHundred} project, we have measured the 3D shapes of the DM halo and stellar (BCG+ICL) distributions as a function of radius, comparing five simulation configurations spanning three DM models ($\sigma_\mathrm{DM}/m = 0$, $0.1$, $0.5\;\mathrm{cm^{2}\,g^{-1}}$), two baryonic physics prescriptions (\textsc{Gadget-X}, \textsc{Gizmo-Simba}), and two resolution levels (3k, 7k). Our main findings are:

\begin{enumerate}

\item \textit{SIDM rounds the DM halo but not the stellar distribution.} DM haloes become significantly rounder in their inner regions with increasing self-interaction cross-section. The stellar (BCG+ICL) distribution retains the same morphology in all models at fixed baryonic physics. In CDM, both shapes are set by the same accretion geometry, which is why they show good correspondence; in SIDM, collisional isotropisation rounds the DM without affecting the still-collisionless stellar component. The axis ratio offsets between the stellar and DM components are larger for early-forming clusters, consistent with the cumulative nature of DM self-interactions.

\item \textit{The DM--stellar shape divergence is confirmed by complementary diagnostics.} The axis ratio difference peaks at $r \sim 0.1\,r_{200}$ in SIDM, the region where collisional isotropisation is strongest but the BCG does not yet dominate, while the major-axis misalignment diverges from the CDM values at similar radii and continues to grow towards the centre. The 3D WOC shows a monotonic decline towards the centre in SIDM that is absent in the CDM models, indicating that a morphological mismatch extends into the innermost regions. A hierarchical Bayesian comparison at $r = 0.1\,r_{200}$ confirms that the decoupling is detected at $>5\sigma$ for $\sigma_\mathrm{DM}/m = 0.5\;\mathrm{cm^{2}\,g^{-1}}$ in all three diagnostics, and remains significant for $\sigma_\mathrm{DM}/m = 0.1\;\mathrm{cm^{2}\,g^{-1}}$ in $\Delta(b/a)$ and $\Delta\mathrm{PA}$.

\item \textit{The signal is concentrated in the inner halo.} The projected WOC analysis of \citet{Yoo2026} found the BCG+ICL to remain a relatively faithful match of DM in SIDM, but the radially resolved diagnostics presented here reveal a strong decoupling concentrated within the inner halo.

\item \textit{The stellar shape is sensitive to baryonic physics, not DM microphysics.} The choice of subgrid model drives clear differences in both the stellar and DM shapes between the \textsc{Gadget-X} and \textsc{Gizmo-Simba} CDM runs, particularly in the BCG-dominated inner regions. By contrast, changing the DM model from CDM to SIDM at fixed baryonic physics produces negligible change in the stellar shape, despite substantially rounding the DM halo.

\end{enumerate}

These results demonstrate that the ICL--DM shape correspondence established in CDM simulations \citep[e.g.][]{Alonso-Asensio2020, Yoo2024, Fernandez2026} does not hold when DM is self-interacting, and that the assumption of a close match between ICL and DM morphology must be revisited in the context of alternative DM models.

\section*{Acknowledgements}

NAH and AF thanks the Leverhulme Trust for support through a Research Leadership Award. NAH and FRP acknowledge support from the UK STFC under grant ST/X000982/1.
WC and GY are partially supported by the Agencia Estatal de Investigación (AEI, Spain)  under project PID2024-156100NB-C21, funded by MCIN/AEI/10.13039/501100011033.
YMB acknowledges support from UK Research and Innovation through a Future Leaders Fellowship (grant agreement MR/X035166/1) and financial support from the Swiss National Science Foundation (SNSF) under project ``Galaxy evolution in the cosmic web'' (200021\_213076). MSF gratefully acknowledges the support of the Alexander von Humboldt Foundation through a Feodor Lynen Research Fellowship. MM acknowledges support from grant RYC2022-036949-I financed by the MICIU/AEI/10.13039/501100011033 and by ESF+, grant PID2024-158845NB-I00 financed by MICIU/AEI/10.13039/501100011033 and by ERDF, EU, and program Unidad de Excelencia Mar\'{i}a de Maeztu CEX2020-001058-M. JY acknowledges support from the Korea Astronomy and Space Science Institute under the R\&D program (Project No. 2026-1-831-03) supervised by the Korea AeroSpace Administration. GM thanks Joseph Butler, Lucas Kimmig, Jesse Golden-Marx, Rhea-Silvia Remus and Ellen Sirks for fruitful discussions.

This work has been made possible by \textit{TheThreeHundred} collaboration, which benefits from financial support of the European Union’s Horizon 2020 Research and Innovation programme under the Marie Skłodowska-Curie grant agreement number 734374, i.e. the LACEGAL project. \textsc{TheThreeHundred} simulations used in this paper have been performed in the MareNostrum Supercomputer at the Barcelona Supercomputing Center, thanks to CPU time granted by the Red Española de Supercomputación. The high-resolution (7k) simulations from \textsc{TheThreeHundred} were performed on multiple Supercomputers: MareNostrum, Finisterrae3, and Cibeles through The Red Española de Supercomputación grants, DIaL3 -- DiRAC Data Intensive service at the University of Leicester through the RAC15 grant, and the Niagara supercomputer at the SciNet HPC Consortium.

This work made use of the following software: \textsc{Matplotlib} \citep{Hunter2007}, \textsc{NumPy} \citep{Harris2020}, \textsc{SciPy} \citep{Virtanen2020}, \textsc{pywoc} \citep{Yoo2022} and \textsc{PyMC} \citep{pymc2023}.

\section*{Data Availability}

\textsc{TheThreeHundred} galaxy clusters sample is available on request following the guidelines of \textit{TheThreeHundred} collaboration, at https://www.the300-project.org.
 


\bibliographystyle{mnras}
\bibliography{paper_mnras} 




\appendix


\bsp	
\label{lastpage}
\end{document}